\documentstyle[12pt]{article}

\begin{document}

\newcommand{\nn}{\noindent}
\newcommand{\non}{\nonumber}
\newcommand{\ga}{\gamma \gamma}
\newcommand{\ra}{\rightarrow }
\newcommand{\SM}{${\cal SM}$}
\newcommand{\MSSM}{${\cal MSSM}$}
\newcommand{\SUSY}{${\cal SUSY}$}
\newcommand{\CP}{${\cal CP}$}
\newcommand{\X}{{\cal H}}

\nn \hspace*{12cm} DESY 92--170 \\
\hspace*{12cm} UdeM--LPN--TH--111 \\

\vspace*{1.5cm}

\centerline{\large{\bf TWO--PHOTON DECAY WIDTHS}}

\vspace*{0.6cm}

\centerline{\large{\bf OF HIGGS PARTICLES}}

\vspace*{1.5cm}

\centerline{\large{\bf A. Djouadi$^1$, M.~Spira$^2$} and \large{\bf
P. M. Zerwas$^2$}}

\vspace*{1.3cm}

\centerline{$^1$ Laboratoire de Physique Nucl\'eaire, Universit\'e de
Montr\'eal,}
\centerline{Case 6128 Succ.~A, H3C 3J7 Montr\'eal PQ, Canada}

\vspace*{0.5cm}

\centerline{$^2$ Deutsches Elektronen--Synchrotron, DESY, D--2000
Hamburg 52, Germany }

\vspace*{1.5cm}

\begin{center}
\parbox{14cm}
{\begin{center} {\bf ABSTRACT} \end{center}
\vspace*{0.3cm}

\nn Two--photon decays of Higgs bosons are important channels for the
search of these particles in the intermediate mass range at the $pp$
colliders LHC and SSC. Dynamical aspects of the Higgs coupling
to two photons can also be studied by means of
the $\ga$ fusion of Higgs particles at high--energy e$^+$e$^-$
linear colliders. Extending earlier analyses which had been restricted
to the Standard Model, we present in this note the QCD radiative
corrections to the $\ga$ decay widths of scalar and pseudoscalar Higgs
particles in multi--doublet extensions of the Higgs sector, as realized
for instance in supersymmetric theories. }

\end{center}

\newpage

\nn The exploration of the electroweak symmetry breaking mechanism
is one of the most important tasks of particle physics.
Within the Standard Model,
the solution of this problem is associated with
the existence of fundamental scalar Higgs particles.
If the Higgs particle in the Standard Model [\SM] is light, with a
mass in the intermediate range below $\sim 200$ GeV, the theory can be
extrapolated perturbatively up to the GUT scale.
However, the hierarchy problem
suggests the supersymmetric extension of the model in this case,
stabilizing the
mass of light Higgs bosons in the background of high--energy
GUT scales. In contrast to the \SM, supersymmetric theories incorporate a
spectrum of Higgs particles $\Phi$, in the minimal version [\MSSM] light and
heavy scalar [\CP--even] particles\footnote{The scalar
particles will generically be denoted by $\X$.} $h,H$,
a pseudoscalar [\CP--odd]
particle $A$, and a pair of charged Higgs particles. \\

\nn The precise prediction of the $\ga$ decay widths of Higgs particles
\cite{R1} is important for two reasons. First, the $\ga$ decay mode
plays a crucial r\^ole for the search of these particles in the lower part
of the intermediate mass range at the $pp$ colliders LHC and SSC \cite{R2}.
Second, the $\ga$ widths can be measured directly by means of $\ga$
fusion at high--energy $e^+e^-$ linear colliders
\cite{R2A,R3A}. Since the
photons couple to Higgs bosons via heavy particle loops, the $\ga$ widths
are sensitive to particle masses, standard and also supersymmetric, well
above the Higgs masses themselves. However, these effects are small if the
particle masses are not generated through the Higgs mechanism.
To exploit this method,
it is therefore mandatory to control properly the radiative corrections
to the $\Phi \ga$ couplings mediated by the standard particles. \\

\nn In this note, we shall present the QCD radiative corrections of the
quark--loop contributions to the $\ga$ widths of the Higgs particles in the
[minimal] supersymmetric extension of the Standard Model, Fig.~1. This demands
the extension of the calculation for the \SM\ Higgs decay \cite{R4,R4a}
in two ways: (i)
Since the decay amplitudes in part of the \SUSY~parameter space
are dominated by $b$--quark loops, the QCD corrections are  to be analyzed for
scalar $h,H$ masses above the fermion--antifermion threshold\footnote{This
problem has first been solved by one of the authors \cite{R5}. The
analysis would agree with the results of a recent preprint \cite{R6} if
the large logarithms in the QCD corrections
were not mapped into the running quark mass
[see Appendix A].}; and (ii) the QCD corrections are to be determined
for pseudoscalar Higgs bosons $A$. \\

\nn The $\ga$ couplings to Higgs bosons are mediated by charged heavy particle
loops [$W$, fermion, chargino, sfermion and charged Higgs boson loops in the
scalar case $h,H$, and fermion and chargino loops
in the pseudoscalar case $A$]. Denoting
the quark amplitudes by $A_Q$ $etc$., the $\ga$ decay rates are given by
\cite{R1,R4}
\begin{eqnarray}
\Gamma(\X \ra \ga) &=& \frac{G_F \alpha^2 M_{\X}^3 }
{128 \sqrt{2} \pi^3}
\left| \sum_Q N_C e_Q^2 g_Q^\X A_Q^\X + g_W^\X A_W^\X
+ \ \cdots \ \right|^2 \\
\Gamma(A \ra \ga) &=& \frac{G_F \alpha^2 M_A^3 }{32 \sqrt{2} \pi^3}
\left| \sum_Q N_C e_Q^2 g_Q^A A_Q^A + \ \cdots \ \right|^2
\end{eqnarray}

\nn where the quark and $W$ amplitudes read at lowest order
\begin{eqnarray}
A_Q^{\X} & \Rightarrow & 2 \tau^{-1} \left[ 1+(1-\tau^{-1})f(\tau)
\ \right] \non \\
A_W^{\X} & \Rightarrow & - \tau^{-1} \left[ 3+2\tau+ 3(2-\tau^{-1})
f(\tau) \
\right] \non \\
A_Q^{A} & \Rightarrow & \tau^{-1} f(\tau)
\end{eqnarray}

\nn The scaling variable is defined as $\tau=M_\Phi^2/4m_i^2$ with $m_i$
denoting the loop mass, and
\begin{eqnarray}
f(\tau) = \left\{ \begin{array}{ll}
\arcsin ^2 \sqrt{\tau} & \ \ \mbox{for $\tau \leq 1$} \\\displaystyle
-\frac{1}{4} \left[ \log \frac{\sqrt{\tau}+\sqrt{\tau-1}}{\sqrt{\tau}
-\sqrt{\tau-1}} -i\pi \right]^2 & \ \ \mbox{for $\tau >1$} \end{array} \right.
\end{eqnarray}

\nn The coefficients $g_i^\Phi$ denote the couplings of the Higgs bosons
[normalized to the \SM~Higgs couplings] to top, bottom quarks and $W$ bosons,
recollected for the sake of convenience in Table~1. \\

\nn \begin{minipage}[l]{50mm}{ {\bf Table~1:} Coefficients $g_i^\Phi$ of the
Higgs couplings to quarks and $W$ bosons in the \MSSM. $\alpha, \beta$ are
mixing angles, tg$\beta=v_2/v_1$ being the ratio of the Higgs vacuum
expectation
values. Numerical values of the coefficients
are presented in Ref.~\cite{R7}, for instance.} \end{minipage}
\hspace*{1.2cm}
\begin{minipage}[r]{85mm}{
\begin{tabular}{|c||cc|c|} \hline
& & & \\
$ \ \Phi \ $ &$ g_t $ & $ g_b $ & $g_W $ \\ & & & \\ \hline \hline
& & & \\
$h$  & \ $ \cos\alpha/ \sin\beta $ \ & \ $ - \sin\alpha/\cos\beta $ \ & \
$ \sin(\beta-\alpha) $ \ \\
& & & \\
$H$  & \ $ \sin\alpha/\sin\beta $ \ & \ $ \cos\alpha/ \cos\beta $ \ & \ $
\cos(\beta-\alpha) $ \ \\
& & & \\
$A$  & \ 1/tg$ \beta $ \ & \ tg$\beta $ \ & \ $ 0 $ \
\\ & & & \\ \hline
\end{tabular}
} \end{minipage}

\vspace*{0.7cm}

\nn The cross section for the $\ga$ fusion of
Higgs bosons is found by folding the parton cross section with the $\ga$
luminosity, see e.g.~\cite{R2A}.
The parton cross section is determined by the $\ga$ width so that
$\langle \sigma (\ga \ra \Phi)\rangle \propto
\Gamma(\Phi \ra \ga)/M_\Phi^3$.
The photon beams are generated either automatically as initial--state
Weizs\"acker--Wil\-li\-ams and beamstrahl photons
in $e^\pm$ collisions,
or they may be generated, in a dedicated effort, by Compton back--scattering
of laser light \cite{R3A}. \\

\nn The QCD corrections to the quark amplitudes can be parametrized as
\begin{eqnarray}
A_Q \ = \ A_Q^{LO} \ \left[ \ 1 \ + \ C \ \frac{\alpha_s}{\pi} \ \right]
\end{eqnarray}
The coefficient $C$ depends on $\tau=M_\Phi^2/4m_Q^2(\mu^2)$ where the
{\it running} quark mass
$m_Q(\mu^2)$ is defined
at the renormalization point $\mu$ which is
taken to be $\mu=M_\Phi/2$ in our analysis; this value is
related \cite{R9a} to the pole mass $m_Q(m_Q^2)=m_Q$ in the on--shell
renormalization scheme by
\begin{equation}
m_Q([M_\Phi/2]^2) = \displaystyle m_Q
    \left[\frac{\alpha_s([M_\Phi/2]^2)}
{\alpha_s(m_Q^2)} \right]^{12/(33-2N_F)} \left\{ 1 +
{\cal O}(\alpha_s^2) \right\}
\end{equation}
\nn The lowest order amplitude $A_Q^{LO}$ is to be evaluated for the same mass
value $m_Q([M_\Phi/2]^2)$.
The choice $\mu =M_\Phi/2$ of the renormalization
point avoids
large logarithms $\log M_\Phi^2/m_Q^2$ in the final results
for Higgs masses
much larger than the quark mass [for details see Appendix A].
$\alpha_s$ is taken at $\mu$ for $\Lambda=200$ MeV. \\

\nn We have evaluated the diagrams of the type shown in Fig.~1 plus the
corresponding counter terms for the running quark mass at
$\mu=M_\Phi/2$.
The 't~Hooft--Veltman $\gamma_5$ prescription
\cite{R10} has been adopted for the dimensional regularization of the
amplitudes. The 5--dimensional Feynman pa\-ra\-me\-ter integrals
have been reduced analytically to 1--dimensional integrals which
have been calculated numerically. \\

\nn The amplitudes $C_\X$ for scalar loops and $C_A$ for
pseudoscalar loops
are shown in Fig.~2a/b as functions of $\tau$. The coefficients are real below
the quark threshold $\tau <1$, and complex above.
Very close to the threshold, within a margin of a few GeV,
the present perturbative
analysis\footnote{By choosing the renormalization point $\mu=M_\Phi/2$
the perturbative threshold $E_{th}=2m_Q(m_Q^2)$ coincides with the
on-mass shell value proper.
A shift between
$M_\Phi/2$ and $M_\Phi$, for instance, affects the
widths very little away from the threshold.}
cannot be applied anymore. [It may
account to some extent for resonance effects in a global way.]
Since $Q\bar{Q}$  pairs cannot form $0^{++}$ states at the threshold,
${\cal I}mC_\X$ vanishes there;
${\cal R}eC_\X$ develops a maximum very close to the threshold. By
contrast, since $Q\bar{Q}$ pairs do form $0^{-+}$ states, the
imaginary part
${\cal I}mC_A$ develops a step which is built--up by the
Coulombic gluon exchange [familiar from the Sommerfeld
singularity of the QCD correction to $Q\bar{Q}$
production in $e^+e^-$ annihilation]; ${\cal R}eC_A$ is
singular at the threshold.
For large $\tau$, both coefficients
approach a common numerical value,
as expected from chiral invariance in this limit. In the
opposite limit, the
QCD corrections can be evaluated analytically,
\begin{center}
$m_Q \gg M_\Phi: \ \ C_{\X} \ \ra \ - 1 \ \ \mbox{\rm and}
\ \ \ C_A \ \ra \ \ 0 $
\end{center}
\nn These results can easily be traced back to the form of the $\ga$ anomaly
in the trace of the energy--momentum tensor \cite{R11} and to the
non-renormalization
of the axial--vector anomaly \cite{R12}, as demonstrated in Appendix B. \\

\nn In Fig.~3a--d the QCD corrected $\ga$ widths for $h,H,A$
Higgs bosons are
displayed in the minimal supersymmetric extension of the Standard Model
[taking into account only quark and $W$ boson loops] for two values
tg$\beta=2.5$ and tg$\beta=20$. While in the first case top loops give a
significant contribution, bottom loops are the dominant component
for large tg$\beta$. The overall QCD corrections are shown in the lower
part
of the figures. The corrections to the widths are small, $\sim {\cal
O}(\alpha_s/ \pi)$ everywhere.
[Artificially large $\delta$ values occur only for specific
large Higgs masses
when the lowest order amplitudes vanish accidentially as a
consequence of the destructive interference between $W$ and quark--loop
amplitudes, see also \cite{R5,R6}.] \\

\nn {\it In conclusion.}
We have calculated the QCD corrections to the decays $h,H,A
\ra \ga$ of the neutral scalar and pseudoscalar
Higgs bosons in the minimal supersymmetric extension of
the Standard Model, that is taken for illustration.
These corrections are well under control across the
physically interesting mass ranges, if the running of the quark masses is
properly taken into account. \\

\vspace*{1.3cm}

\nn {\bf APPENDIX A} \\

\nn The QCD corrected quark amplitude for $\Phi \ra \ga$ may be written in the
general form
\begin{eqnarray}
A_Q= A_Q^{LO} (m_Q) \left\{ 1+ \left[c_1(m_Q)+c_2(m_Q)\log \frac{\mu^2}{m^2_Q}
\right] \frac{\alpha_s}{\pi} +{\cal O}(\alpha_s^2)\ \right\} \nonumber
\end{eqnarray}
where $m_Q \equiv m_Q(\mu^2)$ is the quark mass defined at the renormalization
point $\mu$. The scale in $\alpha_s$ may in principle
be chosen different from $\mu$. For
large $m_Q$, $c_1$ approaches $-1$ for $h,H$, and $0$ for $A$,
while $c_2$
vanishes $\sim 1/m^2_Q$, and no large logarithms appear in this limit.
For $m_Q \ra 0$, however, $c_2$ approaches a finite non-zero value while $c_1$
develops a large logarithm the coefficient of which is given by $c_2$,
\[ \begin{array}{l} c_{2} \ \ra \ 2 \\
c_1 \ \ra \ -c_2 \log(M_\Phi^2/4m_Q^2) + \mbox{const} \end{array} \]
By choosing $\mu=M_\Phi/2$, all large logarithms are eliminated
from the coefficient of $\alpha_s$
and mapped into the effective quark mass of the lowest--order
amplitude. This is reminiscent of the corresponding procedure
for Higgs decays
to fermions pairs \cite{R13}. Had we chosen $\mu=m_Q$ instead \cite{R6},
we would be left with unnaturally large corrections not taking advantage
of renormalization group improvements. \\

\nn A technical remark ought to be added on a subtle problem related to the
't Hooft--Veltman implementation of $\gamma_5$ in the dimensional
regularization scheme which reproduces the
axial--vector anomaly to lowest order \cite{R10} automatically.
The multiplicative renormalization factor of the
scalar $Q\bar{Q}$ current is well--known to be
given by $Z_{\X QQ}=1-Z_2+ \delta m_Q/m_Q$
where $Z_2$ is the wave--function renormalization factor and
$\delta m_Q$ the
additive mass shift. To insure the chiral symmetry relation $\Gamma_5 (p',p)
\ra \gamma_5 \Gamma (p',p)$ in the limit
$m_Q \ra 0$ for the fermionic matrix element of the
pseudoscalar and scalar currents, the renormalization factor of the
pseudoscalar current has to be chosen \cite{R14a}
\[ Z_{AQQ}=Z_{\X QQ} + 8 \alpha_s/(3\pi) \] \\

\nn The additional term supplementing the naive expectation is caused by
spurious anomalous contributions that must be substracted by hand \cite{R14}.

\newpage

\nn {\bf APPENDIX B} \\

\nn For large quark masses compared to the Higgs mass $m_Q \gg M_\Phi$,
low energy theorems can be exploited to calculate the
corrections $C_{\X ,A}$ in this limit. \\

\nn The scalar coupling $\X \ga$ can be derived
\cite{R15} from the
requirement that the matrix element
$\langle 0|\theta_{\mu \mu}|\ga \rangle $ of the trace
of the QCD corrected energy--momentum tensor \cite{R11},
\[ \theta_{\mu \mu} = [1+\delta]m_Q \bar{Q}Q + \frac{1}{4}
\frac{\beta_\alpha}{\alpha} e_Q^2
F_{\mu \nu}F^{\mu \nu} \]
vanish in the low energy limit;
$\delta=2\alpha_s/\pi$ and
$\beta_\alpha=2\alpha^2/\pi\! \cdot\! (1+\alpha_s/\pi+\cdots )$
is the QED/QCD
$\beta$--function.
Since the Higgs bosons are coupled to the mass
operator $(m_Q/v)\:\bar{Q}Q$, the QCD corrected effective Lagrangian
can be
determined immediately,
\[ {\cal L}_{\rm eff}(\X \ga) =\frac{\alpha}{2\pi} \left(\sqrt{2} G_F
\right)^{1/2} e_Q^2 \left( 1- \frac{\alpha_s}{\pi} \right)\X F_{\mu \nu}
F^{\mu \nu}. \]

\nn From the non--renormalization of the anomaly of the
axial--vector current
\cite{R12,R12a} follows the non--renormalization of the $A\ga$ coupling,
\[ \partial_\mu A_\mu = 2m_Q \bar{Q} i \gamma_5 Q +\frac{\alpha}{4\pi}
F_{\mu \nu} \tilde{F}^{\mu \nu} \]
so that the effective Lgrangian
\[ {\cal L}_{\rm eff}(A\ga) = \frac{\alpha}{8\pi} \left(
\sqrt{2}G_F\right)^{1/2} \ A F_{\mu \nu} \tilde{F}^{\mu \nu} \]
is valid to all orders of $\alpha_s$ for the two--photon irreducible
part of the diagrams. \\

\nn {\it Note added.} As a corollary to this result
we observe that the irreducible part of the $Agg$
coupling will not be renormalized either,
and only the reducible diagrams need
be calculated for the QCD corrections to the production process
$gg\rightarrow A$ in the range $m_A\leq 2m_Q$.
Only the virtual corrections are different from
the scalar case, $C\rightarrow \pi^2 + 6 +\frac{1}{6}(33-2N_F)
\log \mu^2/m_H^2$
in the notation of eq.(9) in Ref.\cite{R15}. [A.~D., Madison SSC
Workshop '93; see also R.~Kauffman
and W.~Schaffer, BNL preprint).] \\[1cm]

\nn {\bf ACKNOWLEDGEMENT} \\

\nn We thank B.~Kniehl and J.~H.~K\"uhn for discussions
on quark threshold
effects and renormalization schemes.

\newpage

\nn {\bf FIGURE CAPTIONS}

\begin{description}
\item[Fig.1.] Generic diagram of the QCD radiative corrections to the
Higgs coupling to two photons.

\item[Fig.2.] (a) Real and imaginary part of the radiative corrections
to the quark amplitude of the scalar $\X \ga$ coupling, normalized
to the lowest--order amplitude; (b) the same for the pseudoscalar
$A\ga$ coupling.

\item[Fig.3.] (a) Two--photon widths of the \MSSM~Higgs bosons
$h,H,A$ for $tg\beta=2.5$, and (b) size of the QCD radiative
corrections; (c), (d) the same for $tg\beta=20$.

\end{description}

\end{document}